# Transphonon effects in ultrafast nano-devices

Zhiping Xu[1], Quanshui Zheng[1,*], Qing Jiang[2],
Chi-Chiu Ma[3], Yang Zhao[3], Guanhua Chen[3,*], Huajian Gao[4] and Gexue Ren[1]

[1]*Department of Engineering Mechanics, Tsinghua University, Beijing 100084, China,*

[2]*Department of Mechanical Engineering, University of California, Riverside, USA,*

[3]*Department of Chemistry, the University of Hong Kong, Hong Kong, China,* [4]*Division of Engineering, Brown University, 182 Hope Street, Providence, RI 02912, USA*

**We report a novel phenomenon in carbon nanotube (CNT) based devices, the transphonon effects, which resemble the transonic effects in aerodynamics. It is caused by dissipative resonance of nanotube phonons similar to the radial breathing mode, and subsequent drastic surge of the dragging force on the sliding tube, and multiple phonon barriers are encountered as the intertube sliding velocity reaches critical values. It is found that the transphonon effects can be tuned by applying geometric constraints or varying chirality combinations of the nanotubes.**

It was widely perceived prior to World War II that supersonic flights were prohibited by the sound barrier due to catastrophes that occurred when flying vessels approached the sound speed. Thanks to von Karman and many other pioneers, great progress has been made to understand the transonic effect which had caused drastic reductions of the plane-lifting forces in the catastrophic events. Consequently, supersonic flights have become reality[1]. Figure 1A depicts a US navy aircraft flying at or near the speed of sound. A condensation cloud was generated around the aircraft due to the transonic effect. Now imagine a nanoscale train travelling inside a nanoscale tunnel. Will similar speed barriers be encountered by the superfast nano-train? We have performed a molecular dynamics study of a fasting moving carbon nanotube inside a tunnel made of a larger-diameter nanotube, and have found that a sudden and drastic reduction of the axial sliding velocity always takes place when the velocity enters one of the narrow ranges in the velocity spectrum. Figure 1B shows the inner tube moving towards the reader at a certain speed. The atoms on the inner and outer tubes are displaced significantly from their equilibrium positions, which lead to a huge drag on the moving inner tube. This new phenomenon at the nanoscale bears resemblance to the transonic effect in aerodynamics depicted in Figure 1A, and has vital implications for developing GHz nanotube oscillators based on multiwalled carbon nanotubes (MWNTs) first proposed in 2002[2]. We report in this letter the novel phenomenon and its underlying mechanism.

There has been intense interest[3-9] in investigating energy dissipation mechanisms in CNT-based ultrafast devices via molecular dynamics simulations. These studies have been motivated by the potentials of the nanoscale devices to serve as basic building blocks in next-generation nano-electromechanical systems (NEMS)[10-12]. On the other hand, onset of sliding friction and resonant energy transfer in one-dimensional models, such as the Frenkel-Kontorova chain[13] and the Fermi-Pasta-Ulam lattice[14], have also attracted much recent attention[15-17]. Resonant coupling leading to wide instability windows is found to be responsible for energy dissipation in systems with a quasicontinuous excitation spectrum. An interesting question to pose then is whether such behavior manifests itself in CNT-based devices. This work will give a comprehensive account of the issue.

In MWNT oscillators driven by the van der Waals restoring force between the coaxial cylindrical graphene shells, relative axial speeds between the core and the sheath can reach as high as 1400 m/s[7]. One can envision such a system as nanoscale bullet trains travelling inside a nanoscale tunnels with an extra high speed. If and how sliding energies dissipate into heat constitute some of the intriguing questions. Previous studies show that in the double-walled carbon nanotubes (DWNTs) spanning no less than several nanometers, energy dissipation occurs quickly and axial oscillation vanishes in a few nanoseconds. Internal mechanical modes of the graphene shells are excited by sliding motion[3], and energy conversion is found to be velocity-dependent[7] and nonlinear when the axial speed exceeds a critical value[9]. For instance, Servantie and Gaspard[9] have extracted, from their simulation results, a relation between the friction and the maximum translation velocity of the oscillating tube, and they have reported a transition of the friction-velocity relation from linear to nonlinear at a "critical velocity", suggesting the onset of additional dissipating mechanisms that may "involve the breathing motion of the nanotubes". However, much remains to be understood with regard to dynamic energy-transfer channels in the MWNT oscillators. To this end, we perform a molecular dynamics study of a pair of armchair carbon nanotubes (7,7)@(12,12) via the GROMACS package[18]. In the simulation with time step 1fs, the Dreiding force field[19] is employed to describe the intratube covalent and intertube van der Waals interactions. Periodic boundary conditions with a 5.1 nm supercell are used to simulate an infinite tube length. Furthermore, one carbon atom in the outer tube is fixed axially to retain the relative motion of different graphene shells.

After structural relaxation, various axial velocities $V_0$, ranging from 100 to 2000 m/s, are assigned to the inner tube (the so-called nano-train). The axial intertube speed, plotted in Figure 2A as a function of time, is found to be mostly dependent of the velocity $V_0$. At some critical velocities, such as $V_0 = 1000$, 1100 and 1900 m/s, axial speeds experience sudden and steep drops in amplitude while radial motion of carbon atoms acquires significant excitations. Away from those critical velocities, much less dissipation is found. The intertube van der Waals force has been calculated, which has the same periodicity as $\sqrt{3}a_{C-C} = 0.246$ nm of the DWNT (7,7)@(12,12), where $a_{C-C}$ is the $sp^2$ carbon bond length. Further analysis suggests that the force as a function of the sliding distance $x$ can be written as $f(x) \approx A_1 \sin(2\pi x/\sqrt{3}a_{C-C}) + A_2 \sin(4\pi x/\sqrt{3}a_{C-C})$.

Therefore, as the nano-train travels at a speed V in a nano-tunnel, the intertube van der Waals interactions may ignite resonance if its frequencies $\omega_1 = V/\sqrt{3}\, a_{C-C}$ and $\omega_2 = 2V/\sqrt{3}\, a_{C-C}$, defined as the washboard frequencies (WBF)[15], approach the phonon frequencies of the nanotube systems. As revealed by normal mode analysis, the lowest-energy phonon modes of the (7,7)@(12,12) pair are rigid-body modes with frequencies in the GHz range (such as rigid-body translation and rotation). The radial breathing modes (RBMs) or RBM-like modes have the next lowest phonon frequencies. The RBM-like modes can be further divided into the in-phase and out-of-phase modes with the calculated frequencies 4.4 and 7.6 THz, respectively, for the DWNT (7,7)@(12,12), similar to those reported in the literature[20]. Consequently, the velocities $V_1 \sim 540$ m/s and $V_3 \sim 1080$ m/s ($V_2 \sim 930$ m/s and $V_4 \sim 1870$ m/s) may excite the in-phase (out-of-phase) RBM-like modes with frequencies around 4.4 THz (7.6 THz), which explains those sudden and steep reduction of the intertube axial speeds at $V_0 = 1000$, 1100 and 1900 m/s. The trajectory for $V_0 = 550$ m/s about 5000 ps from the start of the simulation is amplified in the inset of Figure 2A. Therefore, we conclude that the RBM resonance, a primary source that drains the translational energy, is responsible for the sudden drops in the axial speed, similar to the onset of sliding friction in the one-dimensional Frenkel-Kontorova model[15].

In Figure 2B we plot the axial force dragging on the inner tube ($V_0 = 1000$ m/s) versus time. At $t = 4700$ ps the dragging force on the inner tube experiences a sudden surge. Further analysis shows that the surge of the dragging force is caused by large deformation of the tube walls, which is depicted in Fig. 1B. The phenomenon bears resemblance to the transonic effects that had caused drastic reductions of the forces lifting early airplanes in the catastrophic events. Below or above a critical speed range (CSR), the friction is very small while as the speed falls into a CSR the friction is drastically amplified by orders of magnitudes. A CSR serves as a "barrier" for the inner tube to pass through, and we term it as a "phonon barrier". Unlike the transonic phenomenon with one sonic barrier, the inner tube has several phonon barriers. The overall phenomenon is hence named as the transphonon effect due to its resemblance of the transonic effect.

To gain further insight into the catastrophic event, principal components analysis (PCA)[21] has been performed for the atomic trajectories $x_i(t)$, $i = 1, \ldots, 3N$, for the first 500 ps, where $N$ is the total number of atoms. Eigenvalues of the covariance matrix $C_{ij} = \langle(x_i - \langle x_i \rangle)(x_j - \langle x_j \rangle)\rangle$ measure contributions of corresponding principal modes, where $\langle \ldots \rangle$ denotes the time average. And the eigenmodes are average translational, rotational, vibrational modes, or various non-rigid body modes of the trajectory. We plot in Figure 3A the eigenvalues for the leading 50 principal modes with respect to various initial speeds. In all these cases the translational mode (indexed as "1" in Figure 3A) has the largest contribution to the dynamical motion of the system, followed by the intertube rotation (indexed as "2") whose contribution is one to two orders of magnitude smaller. One can see from Figure 3A that contributions from non-rigid body modes (indexed as "3" and above) are negligible when the axial speed V is outside the CSRs. However, for speeds within the CSRs such as $V_0 = 1000$ m/s, many non-rigid modes have been excited. PCA has been carried out for $V_0 = 1000$ m/s at several simulation time intervals to reveal the progressive excitations. The RBM-like modes are excited within the first 100 ps, and reach their saturations at ~500 ps. Translational energies in the inner tube is then dissipated slowly into higher-frequency modes prior to the sudden drop of its sliding speed at 4700 ps. The carbon atoms on the tubes are displaced violently from their equilibrium positions $t = 4700$s, which leads to a huge increase in the dragging force on the sliding tube. This result shows that within the CSRs the RBMs are excited via resonance followed by the excitations of higher frequency modes. Figure 3A reveals the passage of successive excitations of the RBMs and other modes when the sliding speed is within one of the CSRs. Here the resonances are caused by the internal mode couplings, and are thus not the conventional parametric resonances[22]. However, the resonance of the RBM-like modes alone is not sufficient to cause the transphonon phenomenon. The large deformation of the tube walls that follows the resonance of the RBMs is directly responsible to the surge of the dragging force and thus the occurring of the transphonon phenomenon. To examine the excited modes in detail, we have calculated $p_j(t) = x(t) \cdot u_j$, which are projections of the atomic trajectory $x(t)$ onto the leading principal modes $u_j$. As shown in Figure 3B-F, the non-rigid mode that has the largest contribution is the RBM followed by the wavy modes[3]. After they are mostly excited after $t = 500$ ps, the RBMs are enhanced by about one hundred times compared with their earlier amplitudes (Figure 3E). Furthermore, many high energy modes are excited (Figure 3A) due to their couplings with the RBM[15], leading to irreversible energy dissipation. Similar behavior has been observed in other CSRs, such as $V_0 = 1100, 1900$ m/s.

Inspired from the above observations, we propose two means to avoid or reduce the transphonon effects for any given nano-train dimension. One is to reinforce the nano-tunnel to elevate the RBM frequencies to a level higher than the WBFs. For that purpose, we fix in space carbon atoms in the outer tube, and perform the simulation under such a constraint. In Figure 4 we plot its axial speed as a function of time. Due to suppression of its coupling to the RBMs after immobilizing atoms in the outer tube, the axial speed hardly decreases with time. Similar observations have been made previously for DWNT oscillators in which atomic coordinates in the outer tube are fixed to prevent formation of nanotube wavy motion during dynamics simulation[3, 4]. In practice, multiwalled carbon nanotubes can be used to achieve the constraint, our simulation of a triple-walled nanotube (TWNT) (7,7)@(12,12)@(17,17) confirms the suppression due to the confinement of the outmost tube (17,17). It was found that the RBM frequencies, either in-phase or out-of-phase, of DWNTs are nearly independent of tube chiralities and follow approximately a power-law relation $\sim c/D$ on outer tube diameters $D$[20]. This suggests another means to alleviate axial-speed losses by choosing desired combinations (or chiralities) of nanotubes that make up the DWNT to reduce the WBFs to a level lower than the RBM frequencies.

DWNTs (7,7)@(12,12), (12,0)@(21,0) and (11,2)@(12,12), for example, have similar diameters and are all commensurate pairs but with different commensurate lengths 0.246, 0.426 and 1.7 nm, respectively. Thus the WBFs of the second and third DWNTs for a given speed are reduced by factors of 1.7 and 6.9, respectively, compared with those of the first. In Figure 4 we also plot the axial speeds of the nano-train systems (12,0)@(21,0) and (11,2)@(12,12) versus time for $V_0$ = 1000 m/s. As expected, the transphonon phenomenon is not observed in the first 5000 ps. Since $V_0$ = 1700 m/s falls into one of the CSRs of the (12,0)@(21,0), all characteristics of the transphonon effects become evident (see Fig. S1 in Supporting Material). Finally, using smaller DWNTs is another effective means for avoiding the excitation of RBMs because smaller DWNTs have higher RBMs, as implied by the observation of Servantie and Gaspard[9].

Molecular dynamics simulations performed so far are all conducted with an initial temperature 0 K, and energy dissipation that follows heats up the nanotubes to ~5 K if $V_0$ is outside the CSRs, and to ~20 K (30-50 K) if $V_0$ is inside CSRs before (after) the steep drops in speed. Simulations at room temperatures (e.g. 300 K) have also been carried out, and similar transphonon behaviour is found albeit much larger energy losses are observed when the sliding speed is outside the CSRs as compared to those at low temperatures. Furthermore this thermally induced dissipation is further found to be velocity-independent. In addition, we have also carried out simulations with different supercell lengths, e.g. 7.5 and 10 nm, and found that wavy mode depends upon the length[3, 7]. This dependence, as our simulation shows, affects the coupling strengths among the RBM, wavy and axial sliding modes, but leaving unchanged other characteristics of the transphonon phenomenon (e.g. the sudden drop of the sliding speed, the significant deformation of radial motion, and the surge in the dragging force). These results are presented in the supporting material.

To conclude, the transphonon phenomenon is found to be mainly responsible for energy dissipation in an ultrafast nanoscale train-tunnel system composed of MWNTs. In addition to dissipation mechanisms previously identified for DWNT oscillators, such as the rocking motion instability or the end effects[3,7,8], the transphonon behavior is primarily due to resonant coupling of the axial motion to a wide window of phonon modes and the subsequent giant deformation of the tube walls. Multiple phonon barriers are discovered as the intertube velocities reaches resonant values that are determined by the DWNT diameter and commensurate length as well as diameter. The results reported here are believed to have important implications to ultrafast nanodevices in general that possess phonon frequencies prone to velocity-induced resonances of moving parts. Moreover the means to reduce energy losses proposed here will have wide applications in NEMS design.

**Supplementary Information** accompanies the paper on **www.nature.com/nature**.

**Acknowledgements** The work is supported by the National Science Foundation of China through Grants 10172051, 10252001, and 10332020, the Hong Kong Research Grant Councili (NSFC/RGC N HKU 764/05 and HKU 7012/04P), Germany. ZX also thanks Drs. Farid F. Abraham and Markus J. Buehler for their help on simulation.

**Author Information** The authors declare no competing financial interests. Correspondece and requests for materials should be addressed to Z. Q. S (zhengqs@tsinghua.edu.cn) or C. G. (ghc@everest.hku.hk).

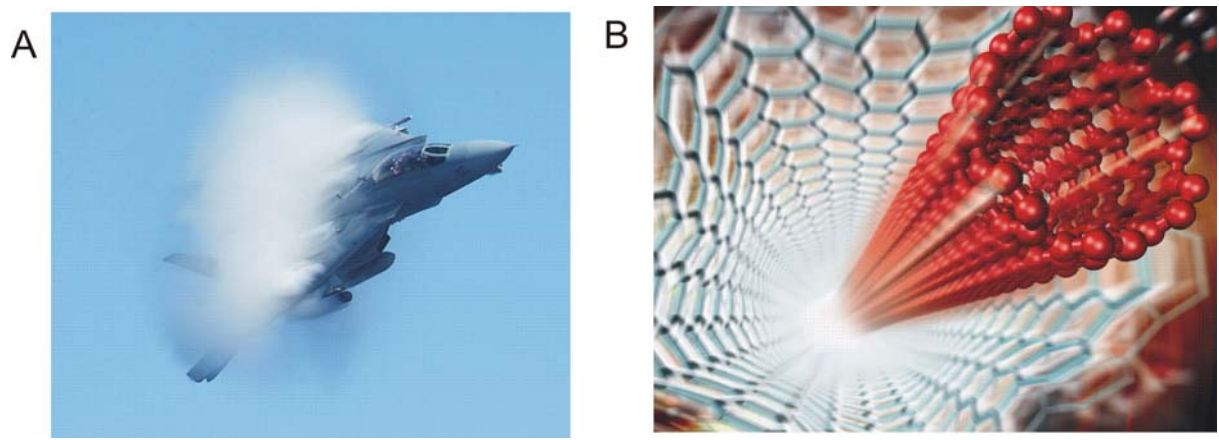

Figure 1

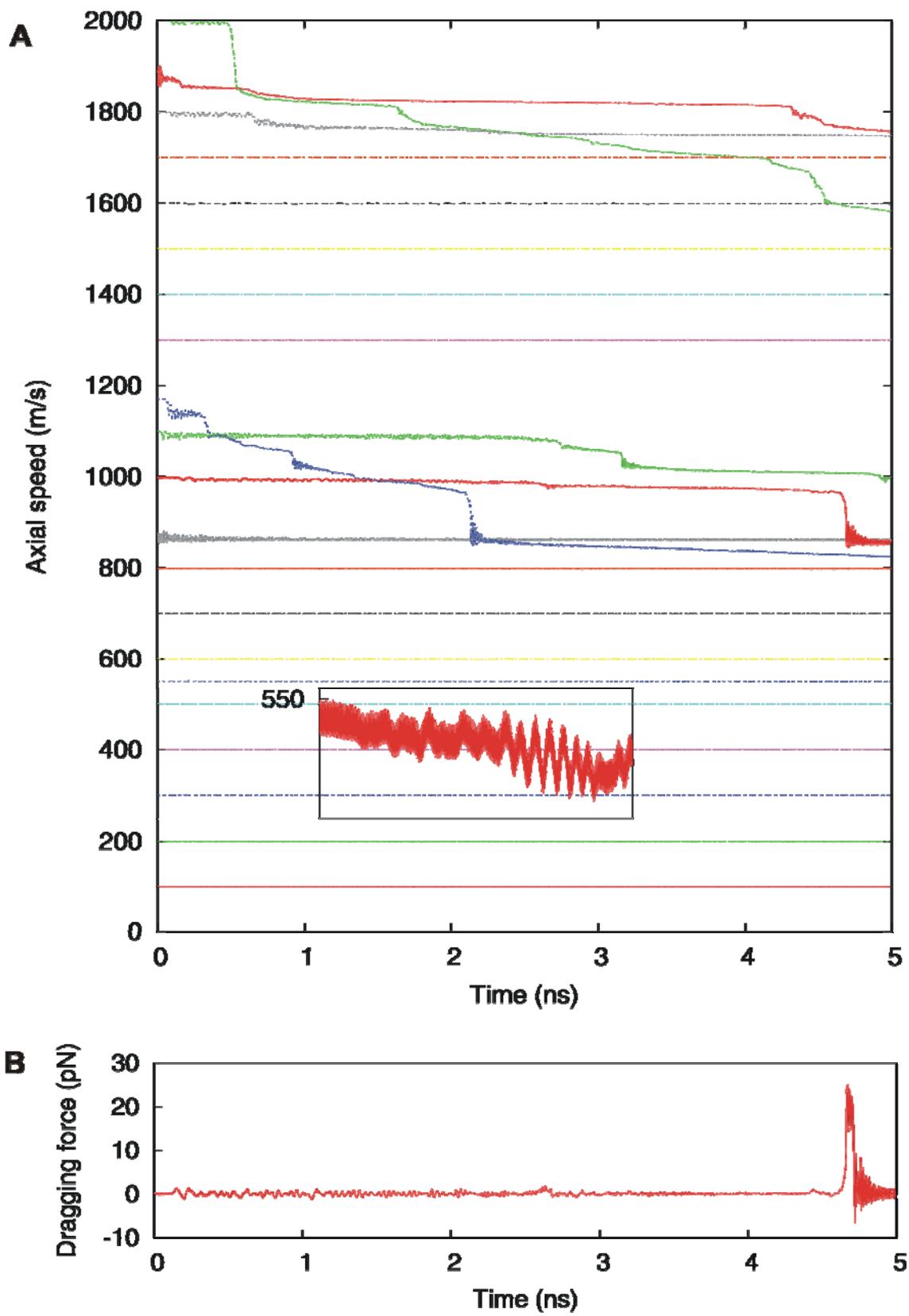

Figure 2

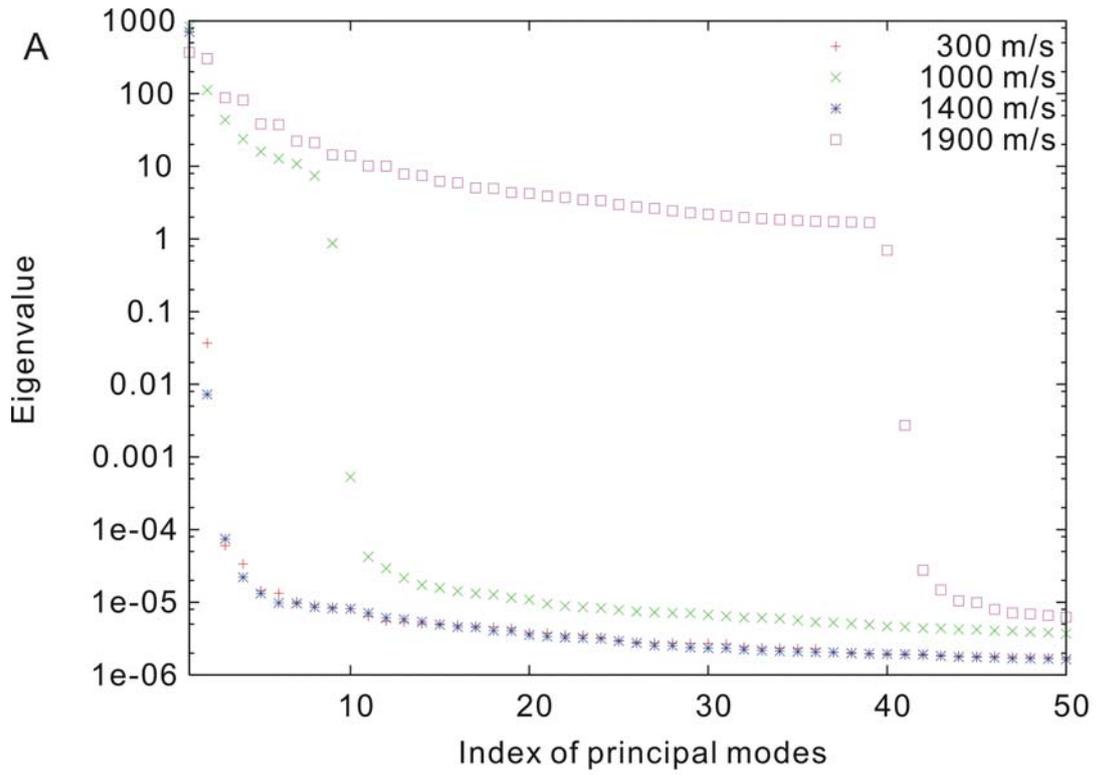
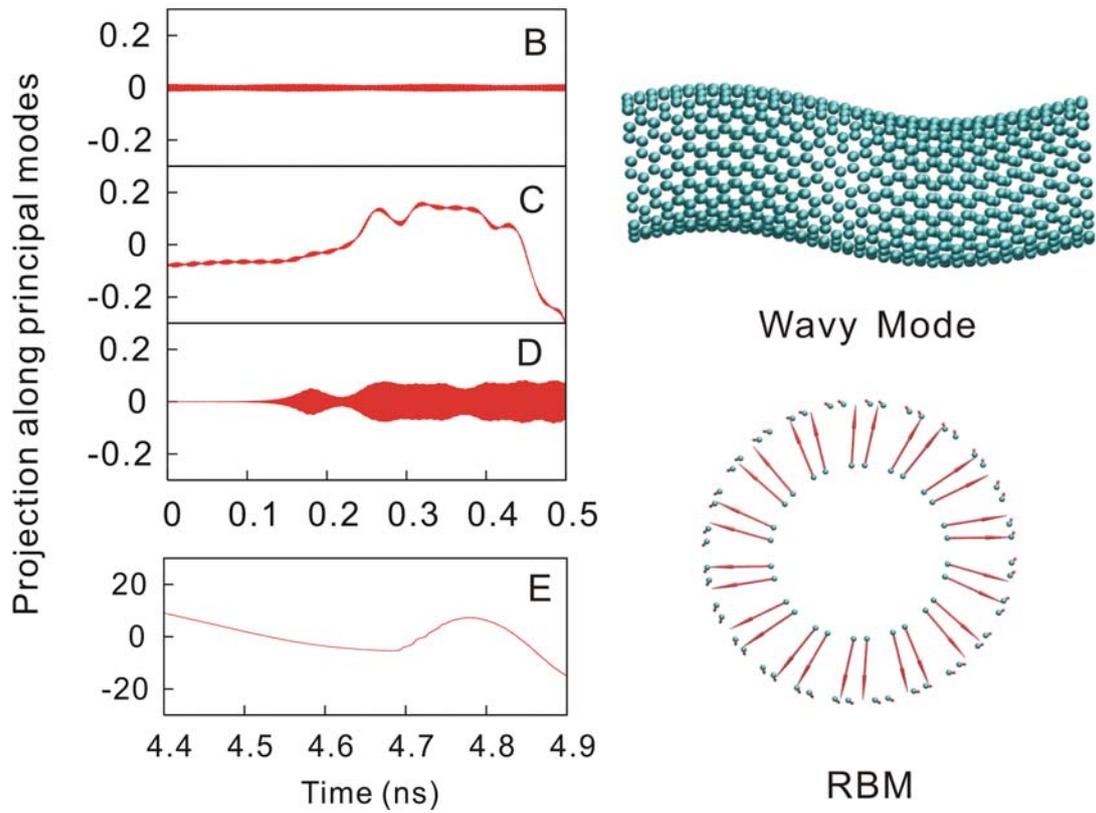

Figure 3

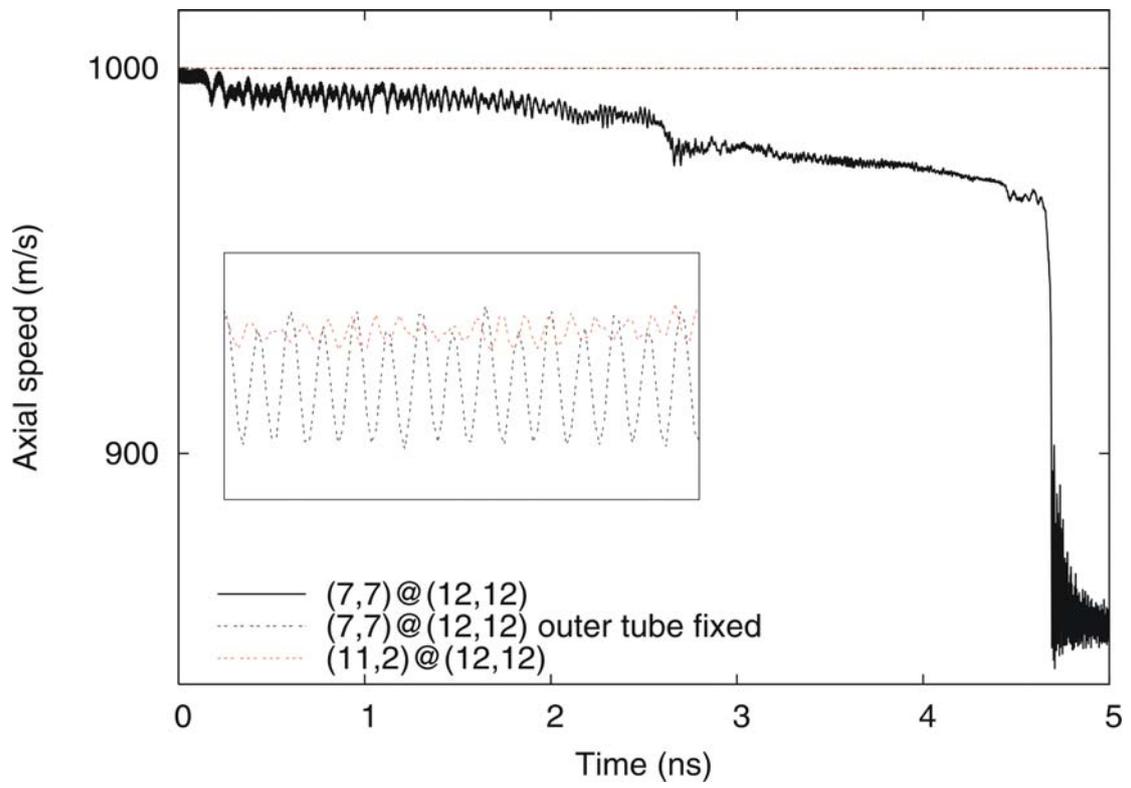

Figure 4

# Figure Captions

Fig. 1 (A) Transonic effect: a photo of an aircraft flying at or near the speed of sound. The white cone is the condensation cloud caused by the transonic effect. (B) An artistic sketch of the transphonon effects of a carbon nanotube moving fast inside a large radius carbon nanotube.

Fig. 2 (A) Time evolution of the axial speed of the nano-train with various initial speed V0 ranging from 100 to 2000 m/s. Sudden drops of the axial speed are found when the speed falls into the CSRs centred at 540 (see the inset), 930, 1080 and 1870 m/s, respectively. Between CSRs the axial speeds have little slow-down within 5 ns. (B) The dragging force on the inner tube versus time. The initial speed of the tube V0 is 1000 m/s. The force is calculated by differentiating the velocities with averaging over 10 ps.

Fig. 3 (A) Eigenvalues of the fifty most dominating principal modes over 0~500 ps. Outside the CSRs ($V_0$ = 300 and 1400 m/s) only few modes such as translation and rotation dominate. In the CSRs ($V_0$ = 1000 and 1900 m/s), the RBM-like modes are excited within the first 100 ps, reach their saturations at ~500 ps. (B-D) Projections of the atomic trajectories onto the principal modes for the same time-interval 0~500 ps. The translational and rotational modes are excluded. (B) shows the contributions from the wavy modes shown in the upper-right for $V_0$ = 300 m/s. (C) is the contribution from the RBM depicted in the lower-right for V0 = 1000 m/s. (D) shows the contributions from wavy or bending modes for $V_0$ = 1000 m/s. (E) is the contribution from the RBM for $V_0$ = 1000 m/s for $t$ = 4400~4900 ps.

Fig. 4 Time evolution of the axial speed for $V_0$ = 1000 m/s (black solid). Two means: (i) elevating RBFs by applying constraints on the outer tube (gray dash) and (ii) reducing WBFs by using chiral tubes (11,2)@(12,12) have effectively avoided the severe energy dissipation due to resonant coupling with radial breathing phonons (red dash).